\documentclass[twoside]{article}

\usepackage{PRIMEarxiv}

\usepackage[utf8]{inputenc}
\usepackage[T1]{fontenc}   
\usepackage{hyperref}    
\usepackage{url}      
\usepackage{booktabs}  
\usepackage{amsfonts}   
\usepackage{nicefrac}   
\usepackage{microtype}     
\usepackage{lipsum}
\usepackage{fancyhdr}     
\usepackage{graphicx}  
\graphicspath{{figs/}} 

\usepackage[table]{xcolor}
\usepackage{amsmath}
\usepackage{amssymb}
\usepackage{amsthm}
\usepackage[round, authoryear]{natbib}
\usepackage{todonotes}
\usepackage{soul}
\usepackage{cancel}
\usepackage{pdflscape}
\usepackage{algorithm}
\usepackage{algpseudocode}
\usepackage{array}
\usepackage{tabularx}
\usepackage{longtable}

\title{A Moment-Based Generalization to\\ Post-Prediction Inference
  \thanks{\textit{\underline{Citation}}: 
  \textbf{Authors. Title. Pages.... DOI:000000/11111.}} 
}

\author{
  {Stephen Salerno} \\
  {Public Health Sciences, Biostatistics} \\
  {Fred Hutchinson Cancer Center} \\
  {Seattle, WA} \\
  {\texttt{ssalerno@fredhutch.org}} \\
   \And
  {Kentaro Hoffman} \\
  {Statistics} \\
  {University of Washington} \\
  {Seattle, WA} \\
  {\texttt{khoffm3@uw.edu}} \\
   \And
  {Awan Afiaz} \\
  {Biostatistics} \\ 
  {University of Washington} \\
  {Public Health Sciences, Biostatistics} \\
  {Fred Hutchinson Cancer Center} \\
  {Seattle, WA} \\
  {\texttt{aafiaz@uw.edu}} \\
   \And
  {Anna Neufeld} \\
  {Mathematics and Statistics} \\ 
  {Williams College} \\
  {Williamstown, MA} \\
  {\texttt{acn2@williams.edu}} \\
    \And
  {Tyler H.~McCormick} \\
  {Statistics} \\ 
  {University of Washington} \\
  {Sociology} \\
  {University of Washington} \\
  {Seattle, WA} \\
  {\texttt{tylermc@uw.edu}} \\
     \And
  {Jeffrey T.~Leek} \\
  {Biostatistics} \\ 
  {University of Washington} \\
  {Public Health Sciences, Biostatistics} \\
  {Fred Hutchinson Cancer Center} \\
  {Seattle, WA} \\
  {\texttt{jtleek@fredhutch.org}} \\
}

\begin{document}

\maketitle

%--- ABSTRACT ----------------------------------------------------------------------

\begin{abstract}
Artificial intelligence (AI) and machine learning (ML) are increasingly used to generate data for downstream analyses, yet naively treating these predictions as true observations can lead to biased results and incorrect inference. Wang et al.~(2020) proposed a method, \textit{post-prediction inference}, which calibrates inference by modeling the relationship between AI/ML-predicted and observed outcomes in a small, gold-standard sample. Since then, several methods have been developed for \textit{inference with predicted data}. We revisit Wang et al.~in light of these recent developments. We reflect on their assumptions and offer a simple extension of their method which relaxes these assumptions. Our extension (1) yields unbiased point estimates under standard conditions and (2) incorporates a simple scaling factor to preserve calibration variability. In extensive simulations, we show that our method maintains nominal Type I error rates, reduces bias, and achieves proper coverage.
\end{abstract}

%--- KEYWORDS ----------------------------------------------------------------------


\section{Introduction}

Artificial intelligence (AI) and machine learning (ML) have enabled researchers to predict, with a remarkable degree of accuracy, data that are either too costly or difficult to measure  \citep{jumper2021highly, rajkomar2018scalable}. However, drawing inference with these predicted outcomes presents significant statistical challenges. Often, researchers treat AI/ML-derived variables as `measured,' leading to biased effect estimates and incorrect conclusions \citep{wang2020methods}. Recently, several methods have been proposed to calibrate inference \citep{wang2020methods, motwani2023revisiting, egami2023using, angelopoulos2023prediction, angelopoulos2023ppi++, miao2023assumption, gan2023prediction, gronsbell2024another} under this class of {\it `inference with predicted data'} (IPD) problems \citep{hoffman2024we}. We focus on one such method, {\it post-prediction inference} (PostPI).

\citet{wang2020methods} provided two important contributions to the field of IPD. They formalized the problem statistically, differentiating the use of AI/ML-derived outcomes in downstream statistical analysis from other related problems. They illustrated how treating these outcomes as observed does not appropriately reflect their uncertainty, both in terms of the true random error in the unobserved outcome and the uncertainty in the upstream AI/ML model, leading to biased point estimates and anti-conservative inference. The authors also introduced a method, PostPI, based on the observation that {\it good} predictions often have a simple relationship with their true, unobserved counterparts. PostPI models this relationship in a small, labeled dataset to recover the associations between the unobserved outcome and features of interest in a larger, unlabeled dataset. In light of recent developments \citep{motwani2023revisiting}, we scrutinize the original assumptions, discuss scenarios in which those assumptions may fail, and propose a moment‐based extension that yields unbiased parameter estimates, Type I error control, and nominal coverage under more standard assumptions.

\section{Post-Prediction Inference Framework and a Proposed Extension}
\label{sec:problems}

\subsection{Notation} 

Let $Y\in\mathbb{R}$ be an outcome that is expensive or time-consuming to measure directly, and let $\boldsymbol{Z}\in\mathbb{R}^{q}$ be a set of readily observed predictors. We are interested in estimating the association between a subset of these predictors, $\boldsymbol{X}\subseteq\boldsymbol{Z} \in \mathbb{R}^p$, and $Y$ in a large {\it unlabeled} sample where $Y$ is not observed, using a smaller {\it labeled} sample, where $Y$ is measured, to calibrate inference. Our data are $\mathcal{L}\cup\mathcal{U}$, where
\[
    \mathcal{L} = \{(Y_i,\boldsymbol{X}_i,\boldsymbol{Z}_i); i = 1, \ldots, n\},\quad \mathcal{U} = \{(\boldsymbol{X}_i,\boldsymbol{Z}_i); i = n + 1, \ldots, n + N\}  
\]
and $N > n$. As the outcomes in the {\it unlabeled} set, $Y_{\mathcal{U}} = \{Y_i\}_{i=n+1}^N$, are missing, we obtain predictions, $f_i = f(\boldsymbol{Z}_i)$, from a `black‐box' model $f:\mathcal{Z}\to\mathcal{Y}$. Notably, we do not have the operating characteristics of $f$, nor access to its training data. This is the case for large language models \citep{fan2024from} or other deep, pre-trained models \citep{jumper2021highly}.

\subsection{Post-Prediction Inference}

PostPI uses the {\it labeled} data to correct inference in the {\it unlabeled} data. To make this concrete, we focus on the PostPI analytic correction for a downstream linear {\it inferential} model, given by
\begin{equation}
\label{eq:trueinf}
    Y = \boldsymbol{X}^\top\beta + \varepsilon,\quad \mathbb{E}[\varepsilon \mid \boldsymbol{X}] = 0.
\end{equation} 
They fit a linear {\it relationship model} in the {\it labeled} set, $\mathcal{L}$,
\begin{equation}
\label{eq:linearrel}
    Y_{\mathcal{L}} = \gamma_0 + \gamma_1 f(\boldsymbol{Z}_{\mathcal{L}}) + \eta,\quad \mathbb{E}[\eta \mid f(\boldsymbol{Z}_{\mathcal{L}})] = 0.
\end{equation}
They then form pseudo‐outcomes, $Y^*_{\mathcal{U}} = \hat{\gamma}_0 + \hat{\gamma}_1 f(\boldsymbol{Z}_{\mathcal{U}})$, in the {\it unlabeled} set, $\mathcal{U}$, where $\hat{\gamma}_0$ and $\hat{\gamma}_1$ are plug-in estimates. Their final estimator regresses $Y^*_{\mathcal{U}}$ on $\boldsymbol{X}_{\mathcal{U}}$, so that
\begin{equation}
\label{eq:origpostpiest}
    \hat{\beta}_{\rm PostPI} = \left(\boldsymbol{X}_{\mathcal{U}}^\top \boldsymbol{X}_{\mathcal{U}}\right)^{-1} \boldsymbol{X}_{\mathcal{U}}^\top Y^*_{\mathcal{U}}
\end{equation}   
However, under \eqref{eq:trueinf} and \eqref{eq:linearrel}, unless $\mathrm{Cov}\left(\boldsymbol{X}, \eta\right) = 0$,
\[
\begin{aligned}
    \beta &= \left[\mathbb{E}(\boldsymbol{X} \boldsymbol{X}^\top)\right]^{-1} \left\{\gamma_1 \mathbb{E}[\boldsymbol{X} f(\boldsymbol{Z})] + \mathbb{E}(\boldsymbol{X} \eta)\right\} \\
    &\neq \gamma_1 \left[\mathbb{E}(\boldsymbol{X} \boldsymbol{X}^\top)\right]^{-1} \mathbb{E}[\boldsymbol{X} f(\boldsymbol{Z})] = \beta_{\rm PostPI},
\end{aligned}
\]  
\textbf{\textit{PostPI assumes}} the prediction error, $\eta$, is uncorrelated with the covariates, $\boldsymbol{X}$. In many realistic settings, $\eta$ will share structure with $\boldsymbol{X}$, so the bias in $\hat{\beta}_{\rm PostPI}$ for $\beta$ is nonzero.

\subsection{Generalized Estimation}

Assuming independent $\mathcal{L}$ and $\mathcal{U}$, we estimate $\hat{\gamma}_0$, $\hat{\gamma}_1$, and $\hat{\eta}_i = Y_i - \hat{\gamma}_0 - \hat{\gamma}_1 f(\boldsymbol{Z}_i)$, from the {\it labeled} set, and covariances from both the {\it labeled} $(\mathcal{L})$ and {\it unlabeled} $(\mathcal{U})$ sets, where we define the following empirical moments:
\begin{gather*}
    \hat{C}^{\mathcal{U}}_{\boldsymbol{X}f} = \frac{1}{N}\sum_{i=n + 1}^{n + N} \boldsymbol{X}_i f_i,\ \hat{C}^{\mathcal{L}}_{\boldsymbol{X}\eta} = \frac{1}{n} \sum_{i = 1}^{n} \boldsymbol{X}_i \hat{\eta}_i,\
    \hat{M}^{\mathcal{U}}_{\boldsymbol{X}\boldsymbol{X}} = \frac{1}{N}\sum_{i=n + 1}^{n + N}\boldsymbol{X}_i \boldsymbol{X}_i^\top. 
\end{gather*}
Then, our extended estimator is given by
\begin{equation}
\label{eq:estcorr}
    \hat{\beta} = \left(\hat{M}_{\boldsymbol{X} \boldsymbol{X}}^{\mathcal{U}}\right)^{-1} \left(\hat{\gamma}_1 \hat{C}_{\boldsymbol{X} f}^{\mathcal{U}} + \hat{C}_{\boldsymbol{X} \eta}^{\mathcal{L}}\right).
\end{equation}
If $\mathbb{E}[\boldsymbol{X}\eta] = 0$, then \eqref{eq:estcorr} reduces to the original PostPI estimator. Otherwise, \eqref{eq:estcorr} extends PostPI to settings where the prediction error is correlated with $\boldsymbol{X}$. For complete derivations, see Appendix A. 

\subsection{Generalized Inference}

To draw inference, Wang et al.~(2020) approximate the conditional variance of $Y_{\mathcal{U}}$ by
\[
    {\rm Var}\left(Y_{\mathcal{U}} \mid \boldsymbol{X}_{\mathcal{U}}\right) \approx \sigma^2_r + \gamma_1^2 \sigma^2_p,
\]
where $\sigma^2_r$ and $\sigma^2_p$ are the error variances from the {\it relationship} model and from regressing $f(\boldsymbol{Z}_{\mathcal{U}})$ on $\boldsymbol{X}_{\mathcal{U}}$, respectively. Using estimates $\hat{\gamma}_1$, $\hat{\sigma}^2_r$, and $\hat{\sigma}^2_p$, their standard error estimator is
\begin{equation}
\label{eq:orig_se}
    SE\left(\hat{\beta} \mid \boldsymbol{X}_{\mathcal{U}}\right) = \sqrt{\left(\boldsymbol{X}_{\mathcal{U}}^\top \boldsymbol{X}_{\mathcal{U}}\right)^{-1}\left(\hat{\sigma}^2_r + \hat{\gamma}_1^2 \hat{\sigma}^2_p\right)}.
\end{equation}
This mirrors linear regression, with a scalar residual variance and inverse Gram matrix, but \textbf{\textit{PostPI assumes}} the residual variance consists of two components, one from the {\it relationship} model ($\hat{\sigma}^2_r$) and one from the `naive' regression ($\hat{\sigma}^2_p$). As the precision matrix scales as $O(1/N)$, the term involving $\hat{\sigma}^2_r$ vanishes as the size of the {\it unlabeled} set ($N$) goes to infinity. Intuitively, the variance contribution of the {\it relationship} model (based on $n$ observations) should not vanish. We propose a simple extension to \eqref{eq:orig_se} based on \eqref{eq:estcorr}. First, denote  
\[
    M = \mathbb{E}(X X^\top),\quad S_{1} = \mathrm{Var}\left[X f(\boldsymbol{Z})\right],\quad S_{2} = \mathrm{Var}\left(X \eta\right).
\]  
We can show that the estimated standard error is  
\begin{equation}
\label{eq:secorr}
    \hat{\mathrm{SE}}(\hat{\beta}) = \sqrt{\frac{1}{N} \left(\hat{M}^{\mathcal{U}}_{\boldsymbol{X}\boldsymbol{X}}\right)^{-1} \left(\hat{\gamma}_1^2 \hat{S}_{1}^{\mathcal{U}} + \frac{N}{n} \hat{S}_{2}^{\mathcal{L}}\right) \left(\hat{M}^{\mathcal{U}}_{\boldsymbol{X}\boldsymbol{X}}\right)^{-1}},
\end{equation} 
with $S_1$ and $S_2$ estimated by their sample analogs in the {\it unlabeled} and {\it labeled} sets, respectively. For scalar $S_1$, $S_2$, and for $n = N$, \eqref{eq:secorr} reduces to \eqref{eq:orig_se}. We form Wald‐type confidence intervals as $c^\top \hat{\beta} \pm z_{1 - \alpha/2} \sqrt{c^\top \mathrm{Var}(\hat{\beta}) c }$.

%--- SECTION 3 ---------------------------------------------------------------------

\section{Simulation Study}
\label{sec:sims}

\subsection{Data Generation}

We simulate independent $(Y, \boldsymbol{Z})$ in {\it training} sets of size $n_t$, {\it labeled} sets of size $n$, and {\it unlabeled} sets of size $N$. We consider three settings that vary the data-generating mechanism and the allocation of the three sets. In Settings 1 and 2, we generate independent standard normal covariates, $Z_1,\ Z_2,\ Z_3,\ Z_4\sim \mathcal{N}(0,1)$, and form the continuous outcome to depend on $\boldsymbol{X} = Z_1$ through a linear term and $\boldsymbol{Z} \setminus Z_1$, through nonlinear terms:
\[
  Y = \beta_1 Z_1 + \tfrac{1}{2} Z_2 + 3Z_3^3 + 4Z_4^2 + \varepsilon, \quad \varepsilon\sim \mathcal{N}(0,4).
\]
We fit a generalized additive model on the training sample, with spline terms for each $Z_j$ to obtain predictions $\hat{f}(\boldsymbol{Z})$, which are then applied to the {\it labeled} and {\it unlabeled} sets. In Setting 1, we let $n_t = n = N = 500$. In Settings 2 and 3, we let $n_t = n = 500$ and $N = 1,000$. In Setting 3, we simulate correlated predictors from a multivariate normal distribution with compound symmetric covariance, $Z_1,\ Z_2 \sim \mathcal{N}_2\left[\mathbf{0}, \left(\begin{smallmatrix} 1 & 0.5\\0.5 & 1 \end{smallmatrix}\right) \right]$, and
\[
    Y = \beta_1 Z_1 + Z_2 + \varepsilon, \quad \varepsilon\sim \mathcal{N}(0,1).
\]
We fit a random forest on the $n_t$ training observations to produce $\hat{f}(\boldsymbol{Z})$, apply it to the {\it labeled} and {\it unlabeled} sets. Across all settings, we vary the effect, $\beta_1 \in \{0,1\}$.

\subsection{Methods}

We compare our extension [\eqref{eq:estcorr} and \eqref{eq:secorr}] to PostPI [\eqref{eq:origpostpiest} and \eqref{eq:orig_se}], three benchmark methods, and four recent IPD methods. For benchmarks, we fit the {\it oracle} regression of $Y_{\mathcal{U}}$ on $\boldsymbol{X}_{\mathcal{U}}$ (note this is not possible in practice), the {\it naive} regression of $f(\boldsymbol{Z}_{\mathcal{U}})$ on $\boldsymbol{X}_{\mathcal{U}}$ (the approach we advocate against), and the {\it classical} regression of $Y_{\mathcal{L}}$ on $\boldsymbol{X}_{\mathcal{L}}$ (a valid, but potentially inefficient, approach). For IPD methods, we consider prediction-powered inference \cite[PPI; see][]{angelopoulos2023prediction}, PPI++ \cite[see][]{angelopoulos2023ppi++}, PSPA \cite[see][]{miao2023assumption}, and Chen \& Chen \cite[see][]{gronsbell2024another}. Our target of inference is $\beta_1$, which characterizes the association between $\boldsymbol{X} = Z_1$ and $Y$ in a simple linear inferential model. We compare the above methods in terms of bias and mean squared error in estimating $\beta_1$, coverage, and type I error/power. All analyses were performed in {\tt R}, version 4.2.2. The code to implement the IPD methods can be found in the {\tt ipd} package \cite[see][]{salerno2024ipd}. The code to reproduce our results can be found at: \href{https://github.com/salernos/postpi\_sample\_size}{https://github.com/salernos/postpi\_sample\_size}.

\subsection{Results}

Our proposal controls type I error and has nominal coverage across all settings (Table \ref{tab:results}). We reproduce the results in \citet{wang2020methods} under their assumptions and with equal set sizes (Setting 1). We also reproduce the result in \cite{motwani2023revisiting} that the naive approach and PostPI do not control type I error or achieve nominal coverage when the sizes of the {\it training} and {\it labeled} sets are small relative to the {\it unlabeled} set (Settings 2 and 3). Further, the naive approach and PostPI are biased when $f(\boldsymbol{Z})$ does not capture $Y \mid \boldsymbol{X}$ (Setting 3). Lastly, methods such as PPI, PPI++, PSPA, and Chen \& Chen, which have theoretical guarantees, do have nominal coverage and type I error rate control.

\begin{table}%[!p]
\caption{Simulation results for each combination of training $\boldsymbol{(n_t)}$, labeled $\boldsymbol{(n)}$, and unlabeled $\boldsymbol{(N)}$ sample sizes, and for true effect sizes $\boldsymbol{\beta_1 = 0}$ and $\boldsymbol{\beta_1 = 1}$. Bold rows indicate the proposed method.}
\label{tab:results}
\tabcolsep=0pt%
\begin{tabular*}{\textwidth}{@{\extracolsep{\fill}}rrrlrrrrrrrrrr@{\extracolsep{\fill}}}
\toprule%
\multicolumn{4}{@{}c@{}}{Setting} & \multicolumn{5}{@{}c@{}}{$\beta_1 = 0$} & \multicolumn{5}{@{}c@{}}{$\beta_1 = 1$} \\
\cline{1-4}\cline{5-9}\cline{10-14}%
$n_t$ & $n$ & $N$ & Method & Bias & MSE$^{1}$ & CI W$^{1}$ & Cov$^{1}$ & T1 Err$^{1,2}$ & Bias & MSE$^{1}$ & CI W$^{1}$ & Cov$^{1}$ & Power$^{2}$ \\ 
\midrule
\multicolumn{14}{l}{$f(\boldsymbol{Z)}$ Captures $Y \mid \boldsymbol{X}$, Equal Set Sizes} \\
\midrule
500 & 500 & 500 & Oracle       &  0.002 & 0.353 & 2.347 & 0.956 & 0.043 &  0.025 & 0.361 & 2.367 & 0.960 & 0.406 \\ 
500 & 500 & 500 & Classical    & -0.031 & 0.367 & 2.363 & 0.944 & 0.055 & -0.010 & 0.357 & 2.365 & 0.956 & 0.390 \\ 
500 & 500 & 500 & Naive        &  0.002 & 0.303 & 1.877 & 0.916 & 0.083 &  0.012 & 0.285 & 1.900 & 0.923 & 0.557 \\
500 & 500 & 500 & PPI          & -0.017 & 0.285 & 2.127 & 0.951 & 0.049 &  0.008 & 0.323 & 2.157 & 0.946 & 0.448 \\  
500 & 500 & 500 & PPI++        & -0.025 & 0.202 & 1.773 & 0.949 & 0.051 &  0.001 & 0.223 & 1.786 & 0.944 & 0.602 \\ 
500 & 500 & 500 & PSPA         & -0.023 & 0.201 & 1.769 & 0.948 & 0.051 & -0.000 & 0.224 & 1.782 & 0.943 & 0.607 \\ 
500 & 500 & 500 & Chen \& Chen & -0.020 & 0.197 & 1.782 & 0.952 & 0.048 & -0.000 & 0.221 & 1.788 & 0.943 & 0.605 \\ 
500 & 500 & 500 & PostPI       &  0.002 & 0.392 & 2.352 & 0.940 & 0.060 &  0.143 & 0.391 & 2.365 & 0.948 & 0.483 \\ 
{\bf 500} & {\bf 500} & {\bf 500} & {\bf Proposed    } & {\bf -0.017} & {\bf 0.344} & {\bf 2.474} & {\bf 0.961} & {\bf 0.038} & {\bf  0.010} & {\bf 0.386} & {\bf 2.511} & {\bf 0.965} & {\bf 0.367} \\ 
\midrule
\multicolumn{14}{l}{$f(\boldsymbol{Z)}$ Captures $Y \mid \boldsymbol{X}$, Unequal Set Sizes} \\
\midrule
500 & 500 & 1000 & Oracle       &  0.002 & 0.179 & 1.675 & 0.951 & 0.049 &  0.006 & 0.174 & 1.676 & 0.951 & 0.668 \\ 
500 & 500 & 1000 & Classical    &  0.003 & 0.348 & 2.367 & 0.960 & 0.040 &  0.032 & 0.354 & 2.357 & 0.951 & 0.406 \\  
500 & 500 & 1000 & Naive        & -0.002 & 0.179 & 1.332 & 0.880 & 0.120 & -0.006 & 0.163 & 1.335 & 0.895 & 0.790 \\ 
500 & 500 & 1000 & PPI          & -0.003 & 0.187 & 1.690 & 0.957 & 0.043 &  0.011 & 0.184 & 1.691 & 0.956 & 0.663 \\ 
500 & 500 & 1000 & PPI++        &  0.001 & 0.162 & 1.573 & 0.947 & 0.052 &  0.014 & 0.162 & 1.570 & 0.955 & 0.720 \\ 
500 & 500 & 1000 & PSPA         &  0.002 & 0.162 & 1.570 & 0.946 & 0.054 &  0.014 & 0.162 & 1.567 & 0.957 & 0.726 \\  
500 & 500 & 1000 & Chen \& Chen & -0.000 & 0.162 & 1.569 & 0.947 & 0.053 &  0.014 & 0.161 & 1.568 & 0.951 & 0.727 \\  
500 & 500 & 1000 & PostPI       & -0.002 & 0.227 & 1.668 & 0.916 & 0.082 &  0.123 & 0.227 & 1.659 & 0.913 & 0.744 \\ 
{\bf 500} & {\bf 500} & {\bf 1000} & {\bf Proposed    } & {\bf -0.005} & {\bf 0.213} & {\bf 1.902} & {\bf 0.964} & {\bf 0.036} & {\bf  0.007} & {\bf 0.209} & {\bf 1.904} & {\bf 0.963} & {\bf 0.557} \\  
\midrule
\multicolumn{14}{l}{$f(\boldsymbol{Z)}$ Does Not Capture $Y \mid \boldsymbol{X}$, Unequal Set Sizes} \\
\midrule
500 & 500 & 1000 & Oracle       & 0.001 & 0.001 & 0.143 & 0.951 & 0.049 & -0.001 & 0.001 & 0.143 & 0.954 & 1.000 \\ 
500 & 500 & 1000 & Classical    & 0.001 & 0.003 & 0.203 & 0.938 & 0.062 & -0.001 & 0.003 & 0.203 & 0.950 & 1.000 \\ 
500 & 500 & 1000 & Naive        & 0.495 & 0.250 & 0.110 & 0.000 & 1.000 &  0.491 & 0.246 & 0.111 & 0.000 & 1.000 \\  
500 & 500 & 1000 & PPI          & 0.001 & 0.005 & 0.276 & 0.949 & 0.051 & -0.001 & 0.005 & 0.280 & 0.946 & 1.000 \\  
500 & 500 & 1000 & PPI++        & 0.001 & 0.003 & 0.202 & 0.941 & 0.059 & -0.001 & 0.003 & 0.202 & 0.946 & 1.000 \\  
500 & 500 & 1000 & PSPA         & 0.000 & 0.003 & 0.201 & 0.942 & 0.058 & -0.001 & 0.003 & 0.202 & 0.947 & 1.000 \\ 
500 & 500 & 1000 & Chen \& Chen & 0.001 & 0.003 & 0.201 & 0.937 & 0.062 & -0.001 & 0.003 & 0.202 & 0.949 & 1.000 \\
500 & 500 & 1000 & PostPI       & 0.149 & 0.024 & 0.202 & 0.152 & 0.847 &  0.181 & 0.038 & 0.232 & 0.177 & 1.000 \\ 
{\bf 500} & {\bf 500} & {\bf 1000} & {\bf Proposed    } & {\bf 0.001} & {\bf 0.006} & {\bf 0.286} & {\bf 0.918} & {\bf 0.082} & {\bf -0.001} & {\bf 0.008} & {\bf 0.387} & {\bf 0.967} & {\bf 1.000} \\ 
\bottomrule
\end{tabular*}
\noindent $^{1}$ MSE: Mean Squared Error; CI W: Confidence Interval Width; Cov: Coverage Probability; T1 Err: Type I Error. \\
\noindent $^{2}$ For $\beta_1 = 0$, we report type I error and for $\beta_1 = 1$ we report power.
\end{table}

\section{Discussion}
\label{sec:disc}

Our goal was to revisit \cite{wang2020methods}, which offered a simple method to correct inference on AI/ML-generated outcomes under potentially restrictive assumptions, shed light on these assumptions, and offer an extension to broaden its applicability. In simulation, PostPI suffered from biases and had standard errors that did not fully account for prediction errors. Our extension more appropriately controls type I error and achieves nominal coverage. In this, we see the classic bias/variance tradeoff, confirming there is no free lunch when using a predicted outcome in place of its true counterpart.

Since \cite{wang2020methods}, several recent works have made notable advances for IPD.~\cite{angelopoulos2023prediction} introduced the notion of a {\it rectifier} to correct inference on the parameter space, rather than the outcome space.~\cite{angelopoulos2023ppi++} refined this method to improve its statistical and computational efficiency, introducing a tuning parameter to adjust the weight placed on the unlabeled predictions.~\cite{miao2023assumption} proposed a method philosophically similar to \cite{angelopoulos2023prediction, angelopoulos2023ppi++}, but using element-wise variance reduction to improve efficiency.~\cite{gan2023prediction} consider the full variance-covariance to weight their estimator. \cite{gronsbell2024another} proposed a method that is valid for missing-at-random outcomes and more efficient. These works all have theoretical guarantees for unbiasedness and efficiency. More recently, \cite{xu2025unified} and \cite{ji2025predictions} have proposed a unifying framework for these, and other IPD methods. 

With the rapid rise of AI/ML methods, directly analyzing predicted outcomes in place of their true counterparts poses an increasingly common, but serious, inferential challenge. Fortunately, there is growing interest in rigorous statistical corrections for IPD. There are still many open problems, primarily those driven by the design and collection of such data in real-world settings. While the goal of this work was to revisit and extend a previously-developed approach in light of more recent findings, future work will bridge additional gaps between IPD and classical inference to inform the design of new studies with constraints on measuring key outcomes. In particular, how the optimality of incorporating feature information from unlabeled data changes with additional budget and cost constraints, or under different sampling strategies. We will further examine how these approaches can be generalized to settings to more complex settings, such as survival endpoints, high-dimensional inference, and data collected under selection bias. It is our hope that the field of IPD will continue to grow in tandem with the current AI/ML revolution, allowing for valid statistical inference under modern data collection and imputation strategies.

%--- ACKNOWLEDGEMENTS ---------------------------------------------------------------

\section*{Competing interests}

JTL reports Coursera courses which generate revenue for both Johns Hopkins University and the Fred Hutchinson Cancer Center. JTL reports co-founding and serving on the board of Synthesize Bio.

\section*{Author contributions statement}

Conceptualization: SS, THM, JTL; Methodology: SS, KH, AA, AN, THM, JTL; Software: SS; Formal analysis: SS; Investigation: SS, KH, AA, AN, THM,  JTL; Visualization: SS; Writing -- Original Draft: SS; Writing -- Review \& Editing: KH, AA, AN, THM, JTL; Supervision: THM, JTL; Funding acquisition: THM, JTL; Project Administration: THM, JTL.

\section*{Acknowledgments}

This work was supported in part by NIH/NIGMS R35 GM144128 and the Fred Hutchinson Cancer Center J.~Orin Edson Foundation Endowed Chair (SS, JTL), NIH/NHGRI U01 HG012039 (JM, QL), NIH/NIMH DP2 MH122405, R01 HD107015, and P2C HD042828 (THM).

%--- REFERENCES ---------------------------------------------------------------------

\newpage

\bibliographystyle{abbrvnat} 

\bibliography{references}  

%=== APPENDICES =====================================================================

\newpage

\appendix

\setcounter{figure}{0}
\setcounter{table}{0}
\setcounter{equation}{0}

\renewcommand\thefigure{\thesection.\arabic{figure}}
\renewcommand\thetable{\thesection.\arabic{table}} 
\renewcommand\theequation{\thesection.\arabic{equation}} 

\section{Additional Analytic Results}

\subsection{Setup}

Focusing on the PostPI analytic correction, which is a special case for linear regression (Section 2.1 of the original paper), Wang et al.~assume we have two independent samples:
\begin{itemize}
    \item A {\bf testing} (i.e., {\it labeled}) set of size $n$: $\{(X_j, Y_j, \hat{Y}_j)\}_{j = 1}^{n}$, and
    \item A {\bf validation} (i.e., {\it unlabeled}) set of size $N$: $\{(X_i, \hat{Y}_i)\}_{i = 1}^{N}$,
\end{itemize}
where $\hat{Y} = \hat{f}(X)$ and $f : \mathcal{X \to \mathcal{Y}}$ is a prediction function trained on a separate training set of size $n_t$ that maps from the space of $X$ to the space of $Y$. The goal is to conduct inference on the relationship between $Y$ and $X$ in the validation set. We assume a true downstream inferential model:  
\[
    Y = X^\top\beta + \varepsilon,\quad \mathbb{E}[\varepsilon \mid X] = 0,
\] 
but note that we only have access to the predicted $\hat{Y}$ in the validation set, and not the true $Y$. Wang et al.~(2020) proceed by:

\begin{enumerate}
    \item Fitting a ``relationship model'' in the testing set:
    \[
        Y_j = \gamma_0 + \gamma_1 \hat{Y}_j + \eta_j,\quad \mathbb{E}[\eta_j \mid \hat{Y}_j] = 0,\quad j = 1, \ldots, n.
    \]
    \item Forming pseudo‐outcomes in the validation set:
    \[
        Y^*_i = \hat{\gamma}_0 + \hat{\gamma}_1 \hat{Y}_i,\quad i = 1, \ldots, N,
    \]
    \item Solving the ordinary least squares equation for $Y^*_i$ on $X_i$ in the validation set. Their final estimator is:
    \[
        \hat{\beta}_{\rm PostPI} = \left(X_u^\top X_u\right)^{-1} X_u^\top Y^*_u
    = \hat{\gamma}_0\left(X_u^\top X_u\right)^{-1}X_u^\top\mathbf1_{N}
      + \hat{\gamma}_1\left(X_u^\top X_u\right)^{-1}X_u^\top\hat{Y}_u,
    \]
where $\hat{\gamma}_0$ and $\hat{\gamma}_1$ are estimates carried forward from the testing set.
\end{enumerate}

\subsection{Error Decomposition}

Assuming the true model
\[
  Y = X^\top\beta + \varepsilon,\quad \mathbb{E}[\varepsilon \mid X] = 0,
\]
and the relationship model
\[
  Y = \gamma_0 + \gamma_1 \hat{Y} + \eta,\quad \mathbb{E}[\eta \mid \hat{Y}] = 0,
\]
then, we have the decomposition 
\[
    \mathrm{Cov}(X,Y) = \mathrm{Cov}\left(X, \gamma_0 + \gamma_1 \hat{Y}\right) + \mathrm{Cov}\left(X, \eta\right) = \gamma_1 \mathrm{Cov}\left(X, \hat{Y}\right) + \mathrm{Cov}(X, \eta).
\]
If we also assume (WLOG) that $X$ is centered, then $\mathbb{E}[X] = 0$. Note that the {\it true} regression parameter satisfies  
\[
   \mathbb{E}[X (Y - X^\top\beta)] = 0 \quad \Rightarrow \quad \mathrm{Cov}(X, Y) = \mathbb{E}[X X^\top] \beta.
\]  
Since $\mathrm{Cov}(X,Y) = \mathbb{E}[X X^\top]\beta$, it follows that
\[
    \beta = \left(\mathbb{E}[X X^\top]\right)^{-1} \left(\gamma_1 \mathbb{E}[X \hat{Y}] + \mathbb{E}[X \eta]\right).
\]
But, we ignore the term $\mathbb{E}[X \eta]$ and solve  
\[
   \gamma_1 \mathbb{E}[X \hat{Y}] = \mathbb{E}[XX^\top] \beta_{\rm PostPI},
\]  
so  
\[
   \beta_{\rm PostPI} = \gamma_1 \left[\mathbb{E}(X X^\top)\right]^{-1} \mathbb{E}[X \hat{Y}].
\]  
So, unless $\mathrm{Cov}\left(X, \eta\right) = 0$, i.e., unless the prediction error, $\eta$, is uncorrelated with $X$, PostPI targets  
\[
   \beta_{\rm PostPI} = \left[\mathbb{E}(X X^\top)\right]^{-1} \gamma_1 \mathbb{E}[X \hat{Y}] \neq  \left[\mathbb{E}(X X^\top)\right]^{-1} \mathbb{E}[X Y] = \beta,
\]  
In most realistic settings, $\eta$ will share structure with $X$, so that bias is nonzero.

\subsection{Corrected Estimator}

We wish to solve the moment condition for the true $\beta$:  
\[
    \mathbb{E}\left[X (Y - X^\top \beta)\right] = 0.
\]  
We can write
\[
    Y = \gamma_0 + \gamma_1 \hat{Y} + \eta,\quad \mathbb{E}[\eta \mid \hat{Y}] = 0,
\]
so
\[
    X (Y - X^\top \beta) = X \left(\gamma_0 + \gamma_1 \hat{Y} - X^\top \beta\right) + X \eta.
\]
Taking expectations and using $\mathbb{E}[X] = 0$:
\[
    0 = \mathbb{E}\left[X (\gamma_0 + \gamma_1 \hat{Y} - X^\top \beta)\right] + \mathbb{E}[X \eta] = \gamma_1 \mathbb{E}[X \hat{Y}] - \mathbb{E}[X X^\top] \beta + \mathbb{E}[X \eta].
\]
Rearranging gives us:
\[
  \mathbb{E}[X X^\top] \beta = \gamma_1 \mathbb{E}[X \hat{Y}] + \mathbb{E}[X \eta].
\]
Still assuming that we have two samples of independent observations,
\begin{itemize}
    \item {\bf Testing}: $\{(X_j, Y_j, \hat{Y}_j)\}_{j = 1}^{n}$,
    \item {\bf Validation}: $\{(X_i, \hat{Y}_i)\}_{i = 1}^{N}$,
\end{itemize}
we note that we can estimate each piece as follows:

\begin{enumerate}
    \item Slope coefficient, $\hat{\gamma}_1$, from the {\bf testing} set,
    \item Residuals, $\hat{\eta}_j = Y_j - \hat{\gamma}_0 - \hat{\gamma}_1 \hat{Y}_j$, from the {\bf testing} set,
    \item Empirical covariances from {\bf testing} and {\bf validation} sets.
\end{enumerate}
Define the following empirical moments, with explicit superscripts to denote the respective sets that they are estimated in:
\[
    \underbrace{\hat{C}^{(u)}_{X\hat{Y}} = \frac{1}{N}\sum_{i=1}^{N} X_i \hat{Y}_i}_{\scriptsize\text{(validation)}},\quad
    \underbrace{\hat{C}^{(l)}_{X\eta} = \frac{1}{n} \sum_{j = 1}^{n} X_j \hat{\eta}_j = \frac{1}{n}\sum_{j=1}^{n}X_j \left(Y_j - \hat{\gamma}_0 - \hat{\gamma}_1 \hat{Y}_j\right)}_{\scriptsize\text{(testing)}},\quad
    \underbrace{\hat{M}^{(u)}_{XX} = \frac{1}{N}\sum_{i=1}^{N}X_i X_i^\top}_{\scriptsize\text{(validation)}}.
\]
Then, our corrected estimator of $\beta$ (under our previous assumptions) is:
\[
    \hat{\beta}_{\rm corr} = \left(\hat{M}_{X X}^{(u)}\right)^{-1} \left(\hat{\gamma}_1 \hat{C}_{X \hat{Y}}^{(u)} + \hat{C}_{X \eta}^{(l)}\right).
\]

\subsection{Proof of Unbiasedness}

Under i.i.d.~sampling and correct model specification,
\[
    \mathbb{E}\left[\hat{C}^{(u)}_{X\hat{Y}}\right] = \mathbb{E}[X \hat{Y}],\quad
    \mathbb{E}\left[\hat{C}^{(l)}_{X\eta}\right] = \mathbb{E}[X \eta],\quad
    \mathbb{E}\left[\hat{M}^{(u)}_{XX}\right] = \mathbb{E}[X X^\top],\quad \text{and}\quad 
    \mathbb{E}[\hat{\gamma}_1]=\gamma_1,
\]
so we have that
\[
  \mathbb{E}[\hat{\beta}_{\rm corr}] = \left(\mathbb{E}[X X^\top]\right)^{-1} \left(\gamma_1 \mathbb{E}[X \hat{Y}] + \mathbb{E}[X \eta]\right) = \beta.
\]

\subsubsection{Variance Estimation and Type I Error Control}

To derive an expression for the variance of our proposed estimator, denote  
\[
    M = \mathbb{E}[X X^{T}],\quad S_{1} = \mathrm{Var}\left(X \hat{Y}\right),\quad S_{2} = \mathrm{Var}\left(X \eta\right),
\]  
and recall our mean‐based estimator,  
\[
    \hat{\beta}_{\rm corr} = M^{-1} \left(\hat{\gamma_1} \mathbb{E}[X\hat{Y}] + \mathbb{E}[X \eta]\right).
\]  
We can show by the delta method and the central limit theorem that  
\[
    \sqrt{N} (\hat{\beta}_{\rm corr} - \beta) \xrightarrow{d} \mathcal{N}\left(0, M^{-1}\left(\gamma_1^2 S_{1} + \frac{N}{n} S_{2}\right)M^{-1}\right).
\]  
So our asymptotic covariance is  
\[
    \mathrm{Var}(\hat{\beta}) \approx \frac{1}{N} M^{-1} \left(\gamma_1^2 S_{1} + \frac{N}{n} S_{2}\right) M^{-1},
\]  
and a plug in estimate of the standard error for the $k$th component is  
\[
    \hat{\mathrm{SE}}(\hat{\beta}_{k}) = \sqrt{\frac{1}{N} \left[\left(\hat{M}^{(u)}_{XX}\right)^{-1} \left(\hat{\gamma}_1^2 \hat{S}_{1}^{(u)} + \frac{N}{n} \hat{S}_{2}^{(l)}\right) \left(\hat{M}^{(u)}_{XX}\right)^{-1}\right]_{kk}}.
\]  
In practice, we estimate the two variance terms, $S_1$ and $S_2$, by their sample analogs from the {\bf validation} and {\bf testing} sets, respectively, so that
\[
    \underbrace{\hat{S}_{1}^{(u)} = \frac{1}{N} \sum_{i = 1}^{N} \left(X_i \hat{Y}_i - \hat{C}_{X \hat{Y}}\right) \left(X_i \hat{Y}_i - \hat{C}_{X \hat{Y}}\right)^\top}_{\scriptsize\text{(validation)}},\quad \underbrace{\hat{S}_{2}^{(l)} = \frac{1}{n}\sum_{j = 1}^{n} \left(X_j \eta_j\right) \left(X_j \eta_j\right)^\top}_{\scriptsize\text{(testing)}},
\]  
with $\hat{C}_{X \hat{Y}} = \frac{1}{N} \sum_{i = 1}^{N} X_i \hat{Y}_i$. We can then form Wald‐type confidence intervals of the form
\[
    c^\top \hat{\beta}_{\rm corr} \pm z_{1 - \alpha/2} \sqrt{c^\top \mathrm{Var}(\hat{\beta}) c },
\]
or we can use a small‐sample $t$‐approximation. Under our assumed models, these have asymptotically correct nominal coverage and type I error control, regardless of the relative sizes of $n$ and $N$.

%=== END ===========================================================================

\end{document}